# Self-supervised phase unwrapping in fringe projection profilometry

Xiaomin Gao,[1] Wanzhong Song,[1,*] Chunqian Tan[1] and Junzhe Lei[1]

[1]*College of Computer Science, Sichuan University, Chengdu, Sichuan, China, 610065*
*\*Corresponding author: songwz@scu.edu.cn*

**Fast-speed and high-accuracy three-dimensional (3D) shape measurement has been the goal all along in fringe projection profilometry (FPP). The dual-frequency temporal phase unwrapping method (DF-TPU) is one of the prominent technologies to achieve this goal. However, the period number of the high-frequency pattern of existing DF-TPU approaches is usually limited by the inevitable phase errors, setting a limit to measurement accuracy. Deep-learning-based phase unwrapping methods for single-camera FPP usually require labeled data for training. In this letter, a novel self-supervised phase unwrapping method for single-camera FPP systems is proposed. The trained network can retrieve the absolute fringe order from one phase map of 64-period and overperform DF-TPU approaches in terms of depth accuracy. Experimental results demonstrate the validation of the proposed method on real scenes of motion blur, isolated objects, low reflectivity, and phase discontinuity.**

Fringe projection profilometry (FPP) is widely used in science, industrial, and medical three-dimensional (3D) measurement due to the advantages of non-contact, dense point clouds and high measurement accuracy [1]. With the growing demands of 3D shape measurement, improving measurement efficiency without sacrificing measurement accuracy has become one of FPP's most necessary but challenging requirements [2].

The typical approach to fast-speed 3D shape measurement with FPP is using high-speed cameras and digital light processing (DLP) technology. But, the employment of high-speed cameras and DLP will result in an increment in the system cost. Therefore, reducing the number of fringe images per 3D reconstruction and algorithmically retrieving the accurate phase information from the limited number of images is critical in the field of FPP. Single-shot FPP is the ideal solution. Fourier transform profilometry (FTP) [3] is one representative single-shot FPP method. However, due to the limitation of bandpass filtering, FTP often fails in scenes of objects with sharp edges, abrupt changes, or non-uniform surface reflectivity. Phase-shifting technology [4] is introduced to FPP to pursue higher measuring accuracy. With phase-shifting technology, one fringe pattern is shifted in multiple (at least three) steps, and each shifted pattern is sequentially projected on the surface of tested objects. Phase-shifting technology can retrieve accurate pixel-wise phase information from deformed fringe images. The phase information obtained using both FTP and phase-shifting is wrapped in $(-\pi, \pi)$ due to the inverse trigonometric operation. For the 3D reconstruction of objects, this wrapped phase should be restored to the continuous phase. This restoring process, known as spatial phase unwrapping (SPU) [5], faces significant challenges when steep objects or multiple isolated objects are present. Therefore, the temporal phase unwrapping (TPU) [6] algorithm is applied to recover the pixel-wise continuous phase. Typical TPU approaches used in FPP systems can be classified into multi-frequency phase-shifting approaches and the ones based on the Gray code [7][8]. Multi-frequency phase-shifting approaches eliminate phase ambiguity by projecting additional groups of phase-shifting fringes with different fringe periods. Gray code-based methods unwrap the wrapped phase by projecting a series of encoded binary Gray-code patterns. For fast-speed high-accuracy 3D measurement applications, the single-camera FPP system with two-frequency phase-shifting is preferred due to the measurement efficiency and system cost [8]. The common two-frequency phase-shifting approach employs two sets of phase-shifting fringe patterns. The high-frequency pattern is for 3D measurement, and the unit-frequency pattern is for phase unwrapping. In two-frequency phase-shifting approaches, the 3D measurement accuracy is determined by the period number of the high-frequency patterns. However, a higher period number will cause fringe order errors during phase unwrapping because of the amplified phase errors. Consequently, for applications with two-frequency phase-shifting approaches, the period number of the high-frequency pattern is usually limited to about 32 or lower [9], resulting in confined measurement accuracy.

In recent years, deep learning (DL) has been successfully applied in FPP [10]. With deep learning, the absolute phase can be retrieved from two-frequency phase maps [11] or one single-frequency phase map with begin and end fringe order [12]. However, most of the works previously reported on DL-based phase unwrapping in FPP need to acquire large amounts of labeled data to train the models, which, even if available, is laborious and requires professional experts. Therefore, there is an urgent requirement for the training phase unwrapping networks with unlabeled datasets. Recently, an untrained deep learning-based phase retrieval method [13] has been proposed for two-camera FPP systems.

Inspired by the works of self-supervised monocular depth estimation [14], this letter proposes a novel self-supervised phase unwrapping method for single-camera FPP systems. The proposed method is a hybrid scheme of physical model and data-driven. During training, a one-period phase map and a 64-period phase map are required. After training, the model can retrieve the absolute fringe order from only one 64-period phase map.

The flowchart of the proposed method is illustrated in Figure 1. The wrapped phase $\varphi$ with the superscript $c$ and $p$ refer to the camera and projector, respectively. $\varphi_l^c$ and $\varphi_h^c$ denote the one-period and 64-period wrapped phase calculated from captured fringe images, respectively. The CNN is used to estimate a soft fringe order $k_{soft}$ for each point of $\varphi_h^c$. Soft fringe order means the fringe order is a floating number instead of a non-negative integer. Multiplying this soft fringe order by $2\pi$ and adding to the wrapped phase produces the absolute phase $\Phi'$, which can be expressed as:

$$\Phi' = \varphi_h^c + 2\pi \times k_{soft}. \quad (1)$$

By virtue of the uniqueness of the absolute phase value on the epipolar line, the corresponding point of a pixel in camera images is determined in projector patterns [15]. At the corresponding point, a one-period wrapped phase value $\varphi_l^p$ and a 64-period wrapped phase value $\varphi_h^p$ are directly retrieved from the projector fringe patterns. When the coordinates of the corresponding points are non-integer, bilinear interpolation is used to improve the accuracy of retrieved phase values. Thus, two synthesized wrapped phase maps, i.e., a one-period synthesized phase map $\hat{\varphi}_l^c$ and a 64-period synthesized phase map $\hat{\varphi}_h^c$ are produced. These two synthesized wrapped phase maps are the same size as the phase map of the camera view. The difference between $\hat{\varphi}_l^c$ and $\varphi_l^c$ provides the first supervisory signal $Loss_1$. The difference between $\hat{\varphi}_h^c$ and $\varphi_h^c$ provides the second supervisory signal $Loss_2$. Because the two supervisory signals are produced with unlabeled data, the training of the CNN is self-supervised. After training, only one 64-period phase map $\varphi_h^c$ is required to be the input, and the trained CNN outputs one soft fringe order map $k_{soft}$. The absolute fringe order $k$ is obtained after a rounding operation on $k_{soft}$. Then, the absolute phase $\Phi$ is obtained and converted into the depth map after phase-to-height mapping.

This letter uses the difference between wrapped phase maps as the self-supervisory signals for phase unwrapping. Two significant issues must be addressed for successfully training the self-supervised phase unwrapping CNNs.

The first issue is eliminating the periodic ambiguity from the wrapped phase map. In FPP, the wrapped phase is extracted from the captured fringe images with the phase-shifting technique, which can be expressed as

$$\varphi = -arctan \frac{\sum_{n=0}^{N} I_n sin(2n\pi/N)}{\sum_{n=0}^{N} I_n cos(2n\pi/N)}, \quad (2)$$

where $N$ is the number of phase-shifting steps, $I_n$ is the $n$-th phase-shifted fringe image described as follows:

$$I_n = A + Bcos[\Phi + 2n\pi/N], \quad (3)$$

here $A$ is the background intensity, and $B$ is the modulation. According to Eq. (2), the wrapped phase is periodically and discontinuously distributed. When training the phase unwrapping network using only the difference between two high-frequency phase maps as the supervised signal, the network tends to estimate a constant value of the fringe order for all pixels. Therefore, an additional supervised signal is introduced, i.e., the difference between two one-period phase maps. The difference between two one-period phase maps is nonperiodic and able to provide a rough but absolute phase for the phase unwrapping of high-frequency phase maps. Finally, the loss function of our CNN is

$$L = w_1 \underbrace{L_1(\varphi_l^c, \hat{\varphi}_l^c)}_{Loss_1} + w_2 \underbrace{L_1(\varphi_h^c, \hat{\varphi}_h^c)}_{Loss_2}, \quad (4)$$

where, $w_1$ and $w_2$ are the weights. We empirically set $w_1 = 1$ and $w_2 = 2$.

The second issue that needs to be solved for self-supervised phase unwrapping is the realization of error back-propagation from the loss function to the first layer of the CNN. Which requires all operations to be differential. If CNN is expected to output a map in which the fringe order at each pixel is a non-negative integer, the *argmax* operator is often used. This operator hinders the error back-propagation. We introduce soft fringe order as,

$$k_{soft} = sigmoid(k_o) \times a, \quad (5)$$

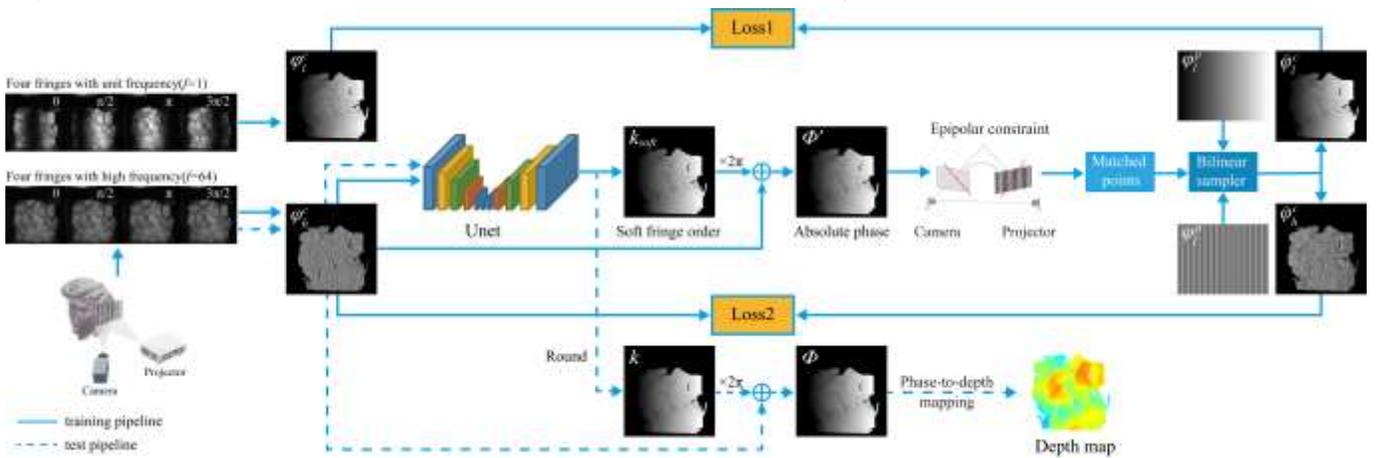

**Fig. 1. Flowchart of the proposed self-supervised phase unwrapping method.**

where, $k_o$ is the output of the last layer of the CNN in Figure 1 and is a one-channel tensor, $a$ represents the period number of high-frequency patterns.

To verify the effectiveness of the proposed method, we employed a handheld FFP system to collect training data. The FFP system comprises one CMOS camera with $1024 \times 1024$ pixels and one DLP projector with a resolution of $684 \times 608$ pixels. Four frequencies, i.e., a period number of 1, 4, 16, and 64, four-step phase-shifting patterns are projected to collect data. The one-period and 64-period phase maps are used to train the network, and the TPU results obtained with the four frequencies are taken as the ground truth. The training dataset contains 7,099 phase maps acquired from twelve dental models; the validation dataset includes 1,385 phase maps from two dental models; the test dataset includes 1,854 phase maps. Among the test dataset, phase maps of two dental models account for 76%, the other 24% of phase maps come from three plaster statue models and one plastic toy model. Data preprocessing includes removing the background and invalid points with a modulation threshold of four, executing morphology operations (erosion followed by dilation) to eliminate noisy data points near the object edges, and removing small areas with less than 1% of the total number of pixels. It is worth noting that more than 80% of the collected data by the handheld FPP system are acquired under non-ideal conditions such as image defocusing and motion blur. Unwrapping these phase maps is challenging for non-DL approaches and DL approaches.

Based on experimental comparisons, UNet [16] is chosen as the backbone CNN. The CNN is trained on NVIDIA Titan RTX using Pytorch 1.8.0. Adam optimizer is adopted with a momentum of 0.9 and a weight decay of 0.0001. The training is divided into two stages; the first stage only uses $Loss_1$ with a learning rate starting from 0.0005 and trains for 50 epochs. The second stage uses the result of the first stage as the pre-trained model, adopts the Loss function in Eq. (4), and then trains for 50 epochs with a learning rate starting from 0.00001. The image size of the input and the output of the CNN is $1024 \times 1024$ pixels.

A series of experiments are conducted to test the performance of the proposed self-supervised phase unwrapping method. In the first experiment, the proposed method is compared with the four-frequency four-step TPU (MF-TPU) and dual-frequency (1, 64) four-step TPU (DF-TPU). The evaluation metric is the RMSE of depth. Figure 2 displays the depth RMSE value distribution on the tested samples. The points included in the depth RMSE calculation are the points with a depth value within the valid range. Over 1,854 test samples, our method produces smaller maximum and average depth RMSE than dual-frequency TPU.

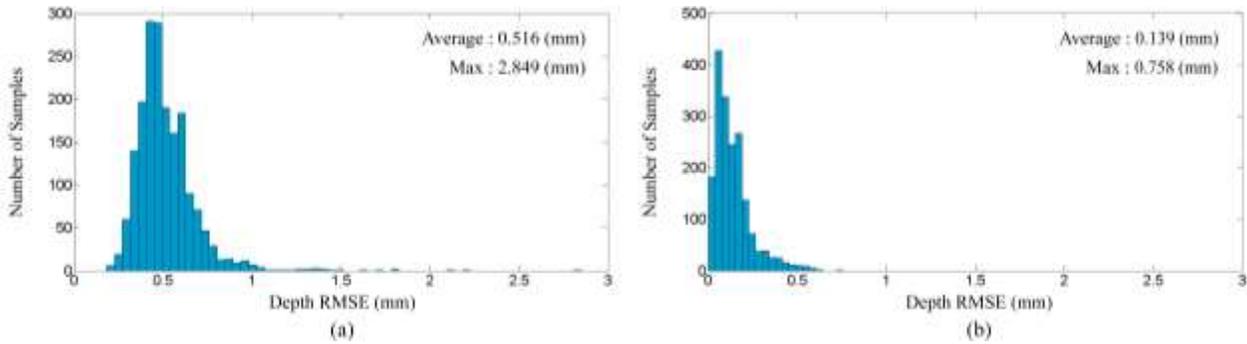

Fig. 2. Distribution of depth RMSE over 1,854 test samples: (a) results of dual-frequency TPU and (b) results of our method.

Figure 3 illustrates the depth maps of some representative samples of the test dataset using the MF-TPU, DF-TPU, and our method. The first row on the left of Figure 3 is an example of motion blur, which results in obvious zig-zag artifacts in the modulation image. The second to fourth rows on the left, the first row on the right, and the second to fourth rows on the right are samples of phase discontinuity, low reflectivity, and isolated objects, respectively. Please note that the training dataset contains only the phase maps of dental models, while the test dataset contains phase maps of models whose shapes are unseen by the network during the training. Our method retrieves the absolute phase of phase maps with motion blur, phase discontinuity, low reflectivity, and isolated objects. In contrast to the conventional DF-TPU, our method significantly improves the accuracy of the continuous phase. The performance of our method is comparable to that of the MF-TPU.

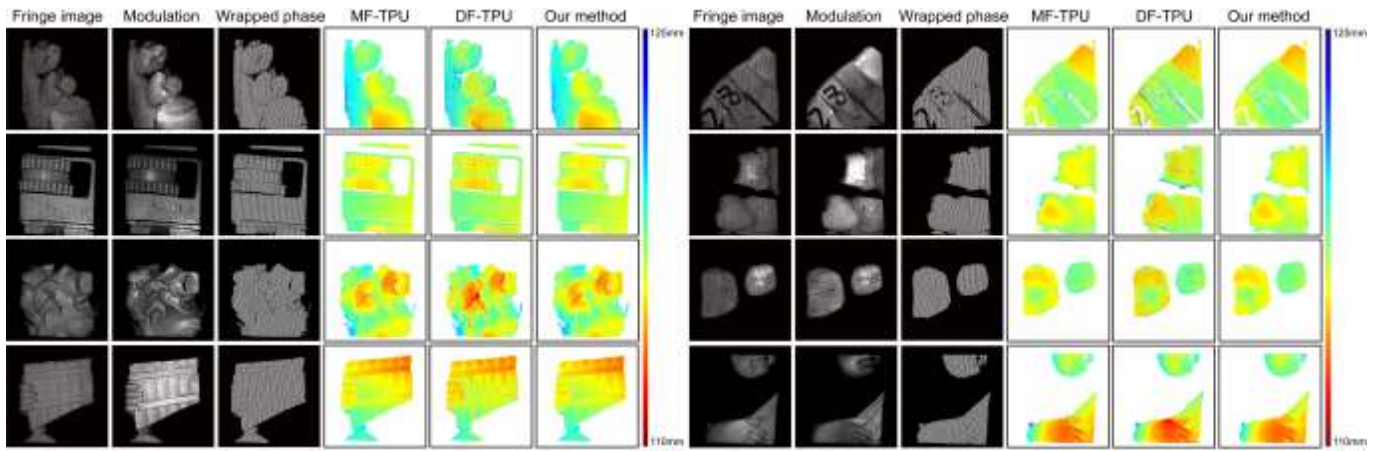

Fig. 3. Reconstructed depth maps. The last three columns of each row are the depth maps using MF-TPU, DF-TPU, and our method.

Next, we conducted ablation experiments to verify the efficiency of the loss function and the network input of the CNN. The ablation experiment includes (#1) only a 64-period phase as the input of the CNN and $Loss_1$ as the loss function; (#2) only 64-period phase as the input of the CNN and $Loss_2$ as the loss function; (#3) a 64-period phase and a one-period phase as the input of the CNN, with $Loss_1$ as the loss function; (#4) a 64-period phase and a one-period phase as the input of the CNN, with $Loss_2$ as the loss function; (#5) a 64-period phase and a one-period phase as the input of the CNN, with $Loss_1$ and $Loss_2$ as the loss function; (#6) the proposed method, only 64-period phase as the input of the CNN, with $Loss_1$ and $Loss_2$ as the loss function. Table 1 depicts the average depth RMSE of the six ablation implementations on the 1,854 test samples.

**Table 1. Ablation study results**

| ID | The input of the CNN | Loss function | Depth RMSE (mm)* |
|---|---|---|---|
| #1 | 64-period phase | $Loss_1$ | 0.470 |
| #2 | 64-period phase | $Loss_2$ | 35.877 |
| #3 | 64-period + one-period phase | $Loss_1$ | 0.517 |
| #4 | 64-period + one-period phase | $Loss_2$ | 35.616 |
| #5 | 64-period + one-period phase | $Loss_1+Loss_2$ | 0.483 |
| #6 | 64-period phase | $Loss_1+Loss_2$ | 0.139 |

Figure 4 illustrates the fringe order maps using the six implementations of the ablation experiment. Compared with the results of MF-TPU, all the depth maps of #1-#5 are erroneous. The proposed method, i.e., #6, produces the best depth map.

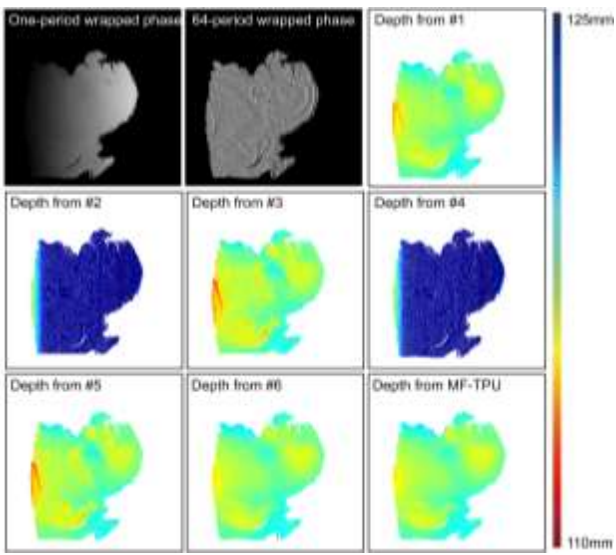

**Fig. 4. Example of the fringe order maps produced by the six implementations of the ablation experiment.**

Finally, to quantify the measurement accuracy of our method, we measured a standard ceramic sphere. Figure 5 shows the ceramic sphere, the fringe image, the wrapped phase, and the depth map using the MF-TPU and our method. The ground truth diameter of the ceramic sphere is 10.0125 mm. The fitting diameter of the ceramic sphere using the MF-TPU and our method is 10.0909 mm and 10.0913 mm, with an error of 0.0784 mm and 0.0788 mm, respectively. This experiment demonstrates the high accuracy of our method.

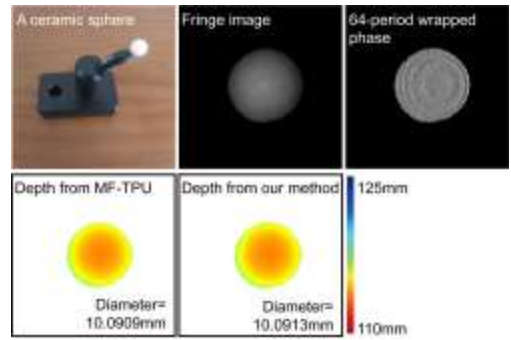

**Fig. 5. Measurement result of a standard ceramic sphere.**

This letter presents a self-supervised phase unwrapping method for single-camera FPP systems. This method can retrieve the absolute fringe order from one high-frequency (64-period) phase map. The qualitative and quantitative performance of the method on real scenes of motion blur, isolated objects, and phase discontinuity is verified. This method could propel the 3D measurement technology.

**Funding.** Research on the Major Special Science and Technology Project of Sichuan Province (2022ZDZX0031); Special Science and Technology Project of Sichuan Province (2021ZYD0104).

**Disclosures.** The authors declare no conflicts of interest.

**Data Availability.** Data underlying the results presented in this letter are not publicly available at this time but may be obtained from the authors upon reasonable request.

**Statement.** The paper is under consideration at Pattern Recognition Letters.